\documentclass[a4paper,fleqn,usenatbib]{mnras}

\usepackage{newtxtext,newtxmath}

\usepackage[T1]{fontenc}
\usepackage{ae,aecompl}

\usepackage{graphicx}	
\usepackage{amsmath}	
\usepackage{amssymb}	

\title[FRBs and Dicke's superradiance]{Explaining fast radio bursts through Dicke's superradiance}

\author[M. Houde et al.]{
Martin Houde,$^{1}$\thanks{E-mail: mhoude2@uwo.ca}
Abhilash Mathews,$^{1}$\thanks{Present address: MIT Plasma Science and Fusion Center, 77 Massachusetts Avenue, NW17-288, Cambridge, MA 02139, USA}
and Fereshteh Rajabi$^{2,1}$
\\
$^{1}$Department of Physics and Astronomy, The University of Western Ontario, 1151 Richmond Street, London, Ontario N6A 3K7, Canada\\
$^{2}$Institute for Quantum Computing, The University of Waterloo, 200 University Ave. West, Waterloo, Ontario N2L 3G1, Canada
}

\date{Accepted 2017 December 8. Received 2017 December 6; in original form 2017 September 30}

\pubyear{2017}

\begin{document}
\label{firstpage}
\pagerange{\pageref{firstpage}--\pageref{lastpage}}
\maketitle

\begin{abstract}
Fast Radio Bursts (FRBs), characterized by strong bursts of radiation intensity at radio wavelengths lasting on the order of a millisecond, have yet to be firmly associated with a family, or families, of astronomical sources. It follows that despite the large number of proposed models no well-defined physical process has been identified to explain this phenomenon. In this paper, we demonstrate how Dicke's superradiance, for which evidence has recently been found in the interstellar medium, can account for the characteristics associated to FRBs. Our analysis and modelling of previously detected FRBs suggest they could originate from regions in many ways similar to those known to harbor masers or megamasers, and result from the coherent radiation emanating from populations of molecules associated with large-scale entangled quantum mechanical states. We estimate this entanglement to involve as many as $\sim10^{30}$ to $\sim10^{32}$ molecules over distances spanning 100 to 1000 AU.
\end{abstract}




\section{Introduction}

At the time of writing this paper, upwards of 20 Fast Radio Bursts (FRBs) have been detected using several radio observatories \citep{Petroff2016}. With the first detection in 2001 \citep{Lorimer2007}, FRBs are seemingly rare but rate calculations for all-sky occurrences predict that one FRB should be detectable on Earth every 10 seconds or so, with the common assumption that they are extragalactic in origin \citep{Thornton2013,Luan2014,Champion2016}. Although no two FRBs are identical, they share a few remarkable similarities. Despite their large luminosity, the small duration of FRBs (on the order of a millisecond) imply that candidate sources are both compact (on the order of 1000 km or less for non-relativistic regimes \citep{Lorimer2007}) and rely on an energetic process, possibly linked to the emission of coherent radiation \citep{Thornton2013,Luan2014}. 

FRBs are broadband signals often spanning tens or hundreds of MHz detected over a range of radio frequencies approximately centred around 1 GHz, while a given FRB can be detected at more than one frequency \citep{Spitler2014,Spitler2016,Scholz2016,Law2017,Gajjar2017}. Some FRBs can show high levels of polarization \citep{Ravi2016}, while they invariably display large dispersion measures ($\mathrm{DM}$) that favor an extragalactic origin interpretation. The measured $\mathrm{DM}$ is responsible for a systematic delay $\delta t \propto \mathrm{DM}\cdot\nu^{-2}$, with $\nu$ the frequency, observed across FRB spectra in a manner consistent with propagation through a cold diffuse plasma. Furthermore, the temporal width $W$ of an FRB pulse is often seen to vary across the spectrum approximately as $W \propto \nu^{-4}$. This pulse broadening behavior and the accompanying exponential tail are as would be expected from multi-path propagation plasma delays \citep{Lorimer2007,Thornton2013,Katz2016}. However, it has been noted that propagation through the intergalactic medium is unlikely to account for this scatter broadening effect, with the implication that most of the measured $\mathrm{DM}$s are probably produced at the location of FRB sources \citep{Luan2014}. The lack of apparent correlation between $W$ and $\mathrm{DM}$ is also consistent with this interpretation \citep{Katz2016}.  

In accordance with the detection of pulses across the full observed frequency range, the whole width of an FRB spectrum (i.e., tens to hundreds of MHz at 1.4 GHz) is not associated to the millisecond duration of its pulse. Instead, as was found by \citet{Ravi2016} for FRB 150807, the FRB spectrum exhibit a decorrelation on a frequency scale consisting of only a small fraction of the total spectral width (estimated to be on the order of approximately 100 kHz for FRB 150807). Still, such frequency scale implies that the underlying temporal structure of the FRB could be characterized by a timescale that is smaller than that of the pulse. As we will see below, this is consistent with the nature of superradiance (SR) bursts as well.     
It is also interesting to note that one FRB (i.e., FRB 121102) has been observed to repeat, albeit with no obvious periodicity \citep{Spitler2014,Spitler2016,Scholz2016,Tendulkar2017,Law2017,Gajjar2017}. The existence of this source is important for discarding FRB models based on cataclysmic events. FRB 121102 also stands out by the fact that it was detected in three frequency bands (i.e., around 1.4 GHz, 2.5-3.5 GHz, and 4-8 GHz) at separate observatories, while the measured spectra are highly variable and do not show the signature of the aforementioned multi-path scattering \citep{Scholz2016}. 

While the astronomical origin of FRBs is now widely accepted \citep{Caleb2017}, a firm association with specific types of sources or physical phenomena has yet to be accomplished. In this paper, we investigate whether Dicke's SR, for which evidence has recently been found in the interstellar medium, can account for the observed characteristics of FRBs and therefore provide a viable physical model for their existence \citep{Mathews2017}.

We start by providing a brief introduction to and summary of the main features of SR in Sec. \ref{sec:SR summary}, then describe in Sec. \ref{sec:methodology} our methodology for the SR modelling of previously published data for FRB 110220, FRB 150418 and FRB 121102 presented and discussed in Sec. \ref{sec:discussion}, while a short conclusion follows in Sec. \ref{sec:Conclusion}. Finally, a detailed summary of the main aspects of SR theory and modelling will be found in the Appendix at the end.

\section{Dicke's superradiance}\label{sec:SR summary}

Since its introduction by \citet{Dicke1954} and its first experimental verification nearly 20 years later \citep{Skribanowitz1973} SR has become, and still is, a very active field of research in the quantum optics community. Although it has remained unnoticed by the astrophysics community until very recently \citep{Rajabi2016A}, evidence has since been uncovered for the occurrence of SR in the interstellar medium (ISM;  \citealt{Rajabi2016B,Rajabi2016Thesis,Rajabi2017}). 

In a nutshell, SR is a quantum mechanical and coherent behaviour between molecules (or atoms) acting as a single quantum mechanical unit as opposed to independent entities. This entanglement is responsible for the emission of powerful pulses or bursts of radiation on short timescales. For example, an ensemble of $N$ molecules under normal circumstances would (for a given transition) each independently spontaneously emit a photon at a wavelength $\lambda$ over a timescale $\tau_\mathrm{sp}$; although under SR conditions will do so cooperatively over a much smaller interval governed by the characteristic timescale 

\begin{equation}
    T_\mathrm{R}=\tau_\mathrm{sp}\frac{8\pi}{3nL\lambda^2}  \label{TR},
\end{equation}

\noindent where the SR sample was assumed to be a thin cylinder of length $L\gg\lambda$ and $n$ is the inverted molecular density (i.e., $n=N/\left(AL\right)$, with $A$ the cross-section of the cylinder). The building up of the aforementioned entanglement in the system will also lead to a time delay 

\begin{equation}
    \tau_\mathrm{D}\simeq\frac{T_\mathrm{R}}{4}\left|\ln\left(\pi\sqrt{N}\right)\right|^2 \label{tau_D} 
\end{equation}

\noindent before the emission of the intensity burst, a behaviour characteristic of SR. It is seen from Eq. \ref{TR} that the timescale of cooperative spontaneous emission varies (decreases) as $N^{-1}$, while the intensity of radiation can be shown to increase with $N^2$ (see Appendix \ref{sec:modelling}). This is in stark contrast with the linear dependency in $N$ for the intensity of a group of independent radiators \citep{Dicke1954,Dicke1964,Gross1982,Benedict1996}. SR therefore leads to coherent radiation, which is also highly focused over a small angular beam (hence the cylindrical shape for the SR sample).

A few basic requirements are needed for SR to take place. More precisely, the sample of entangled molecules must be inverted, velocity coherence between them must be strong enough to bring about their interaction, and so-called dephasing and non-coherent relaxation effects (e.g., collisions) must occur on a timescale $T^\prime> \tau_\mathrm{D}$. Given the first two conditions, it is not surprising that evidence for SR in astrophysics was found in regions known to harbor astronomical masers, with molecular lines known for their association with masers \citep{Rajabi2016B,Rajabi2017}. It can also be shown from the dephasing condition and Eqs. \ref{TR}-\ref{tau_D} that SR requires a critical threshold for the inverted column density

\begin{equation}
    \left(nL\right)_\mathrm{crit}\approx \frac{2\pi}{3\lambda^2}\frac{\tau_\mathrm{sp}}{T^\prime}\left|\ln\left(\pi\sqrt{N}\right)\right|^2 \label{nL_crit}
\end{equation}

\noindent to be exceeded \citep{Gross1976,Rajabi2016B}. SR cannot occur whenever $nL<\left(nL\right)_\mathrm{crit}$, only astronomical masers in a steady-state regime could exist under these less constraining conditions \citep{Rajabi2016B}.   

As evidence for SR in the ISM has so far only been obtained for the OH 1612 MHz, CH$_3$OH 6.7 GHz and H$_2$O 22 GHz spectral lines, and that most FRBs have been detected at about 1 GHz, we will choose the OH 1612 MHz for our analysis. We should keep in mind, however, that similar results could also be obtained with other lines known (or yet to be discovered) to exhibit a population inversion. Further possible SR transitions include, but are not limited to, the 21 cm hydrogen line \citep{Rajabi2016A}, the other three OH maser lines at approximately 1.7 GHz, as well as those at 4.7, and 6.0 GHz, or the CH 3.3 GHz and H$_2$CO 4.8 GHz maser lines \citep{Gray2012}. Evidently, each line would have an associated redshift to match the frequency at which the data were obtained.

We note that although the conditions necessary for the maser action (i.e., population inversion and velocity coherence) are also required for SR (but not sufficient since Eq. \ref{nL_crit} must also be satisfied), the two phenomena are fundamentally different and should not be confused. More precisely, an astronomical (mirrorless) maser can be described as a collective process where stimulated emission events happening at distinct locations in space are responsible for the emission of non-coherent radiation, with an intensity scaling linearly with the number of molecules $N$ involved in the process. SR, on the other hand, cannot be analyzed or thought of as a succession of uncorrelated radiative events. The entangled nature of the underlying quantum mechanical states characterizing a SR sample emphasizes the fact that the ensemble of molecules evolves as a single system. As was previously mentioned, the radiation emanating from a SR sample is coherent in nature and therefore leads to an intensity that scales as $N^2$. The existence of a delay time before the emergence of radiation that is characteristic of SR is not observed for astronomical masers or other types of non-coherent radiation (see Appendix \ref{sec:modelling}).

\section{Methodology}\label{sec:methodology}

As previously mentioned, we provide in Appendix \ref{sec:modelling} a summary of SR theory and the resulting equations entering the models to be presented in Section \ref{sec:discussion}. The SR systems considered in our analysis are so-called ``large samples'' of cylindrical geometry with a length much larger than the wavelength of radiation (i.e., $L \gg \lambda$; see Appendix \ref{sec:large-sample}).  

All SR models were obtained using the sine-Gordon solution described with Eqs. \ref{eq:N-sg}-\ref{eq:TR-large}. We followed the procedure put forth by \citet{Rajabi2017} where for a given data set several realizations of cylindrical SR samples were combined to produce a final SR burst, which was fitted to the dedispersed FRB data. The amplitude of the SR curve was scaled to the data and three parameters were adjusted for the fit: the mean SR characteristic timescale $\left<T_\mathrm{R}\right>$ (or equivalently the mean inverted column density $\left<nL\right>$, through Eq. \ref{TR}), the width $\sigma_{T_\mathrm{R}}$ (i.e., standard deviation) of the Gaussian distribution of $T_\mathrm{R}$ values, and an appropriate dephasing/non-coherent relaxation timescale $T^\prime$. The radii of the cylindrical samples were set by imposing a Fresnel number of unity (i.e., the cross-section of the cylinder is given by $A=\lambda L$) to minimize diffraction and transverse effects not included in the one-dimensional model \citep{Gross1982}. This also ensures that phase coherence is maintained across their length  (see \citealt{Rajabi2017} for more details).

The data of FRB 110220 were taken from \citet{Thornton2013}, while those for FRB 150418 were previously published in \citet{Keane2016} and kindly provided for our analysis by the authors. 

\section{Results and discussion}\label{sec:discussion}

\begin{center}
   \begin{figure}
        \includegraphics[width=\columnwidth]{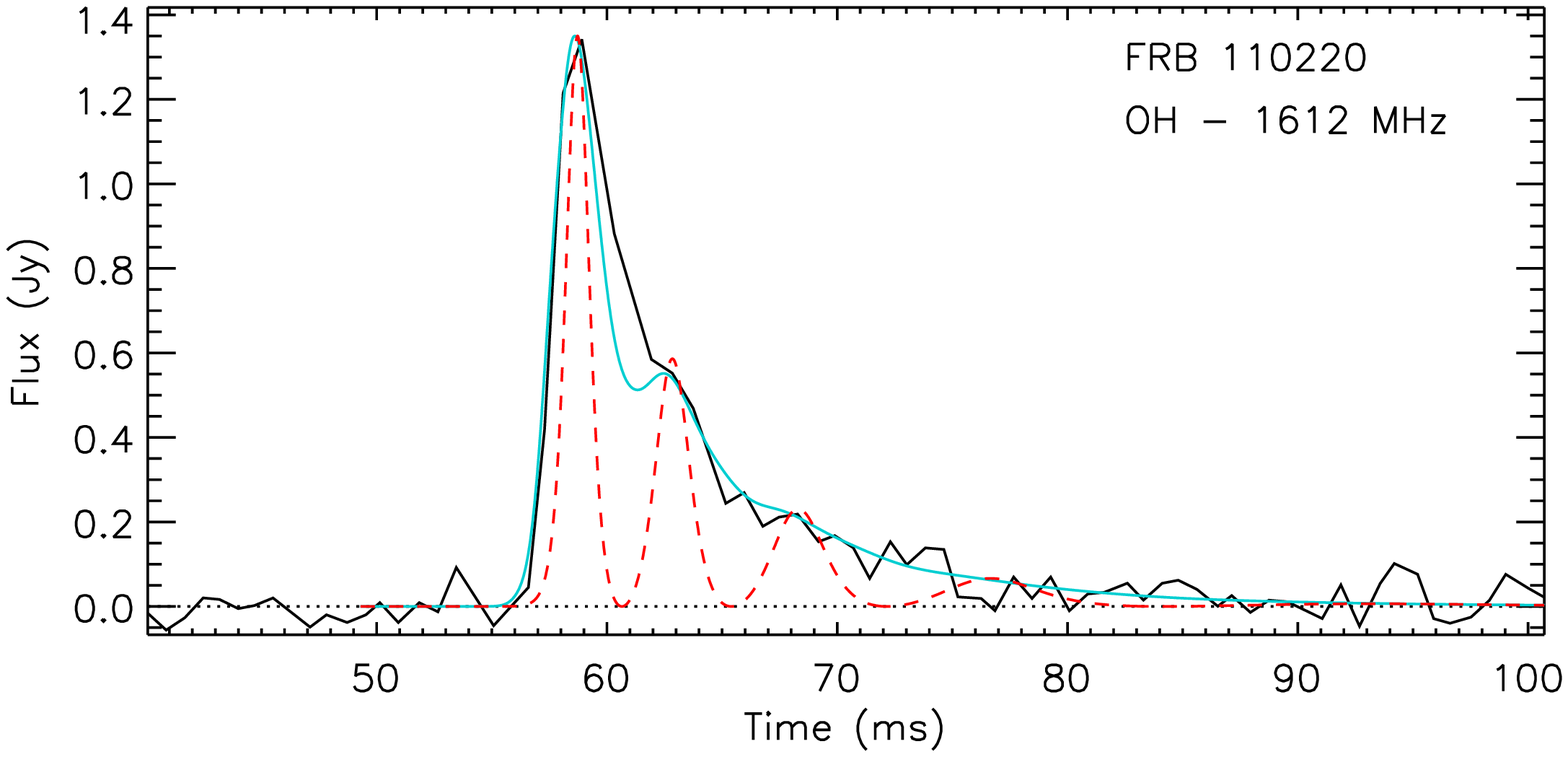}
        \caption{SR model for FRB 110220 \citep{Thornton2013}. The black and cyan solid curves trace, respectively, the data and the resulting fits, while the broken red curve is for a single SR sample with $\left<T_\mathrm{R}\right>=20\,\mu\mathrm{sec}$. The two other model parameters are $\sigma_{T_\mathrm{R}}=0.07\left<T_\mathrm{R}\right>$ and $T^\prime=850\left<T_\mathrm{R}\right>$. The mean inverted column density resulting from the fit is $\left<nL\right>=9.5\times10^{13}$ cm$^{-2}$, which for $\left<n\right>=0.1$ cm$^{-3}$ yields $\left<L\right>=9.5\times10^{14}$ cm.}
        \label{fig:FRB110220}
    \end{figure}
\end{center}

\begin{center}
   \begin{figure}
        \includegraphics[width=\columnwidth]{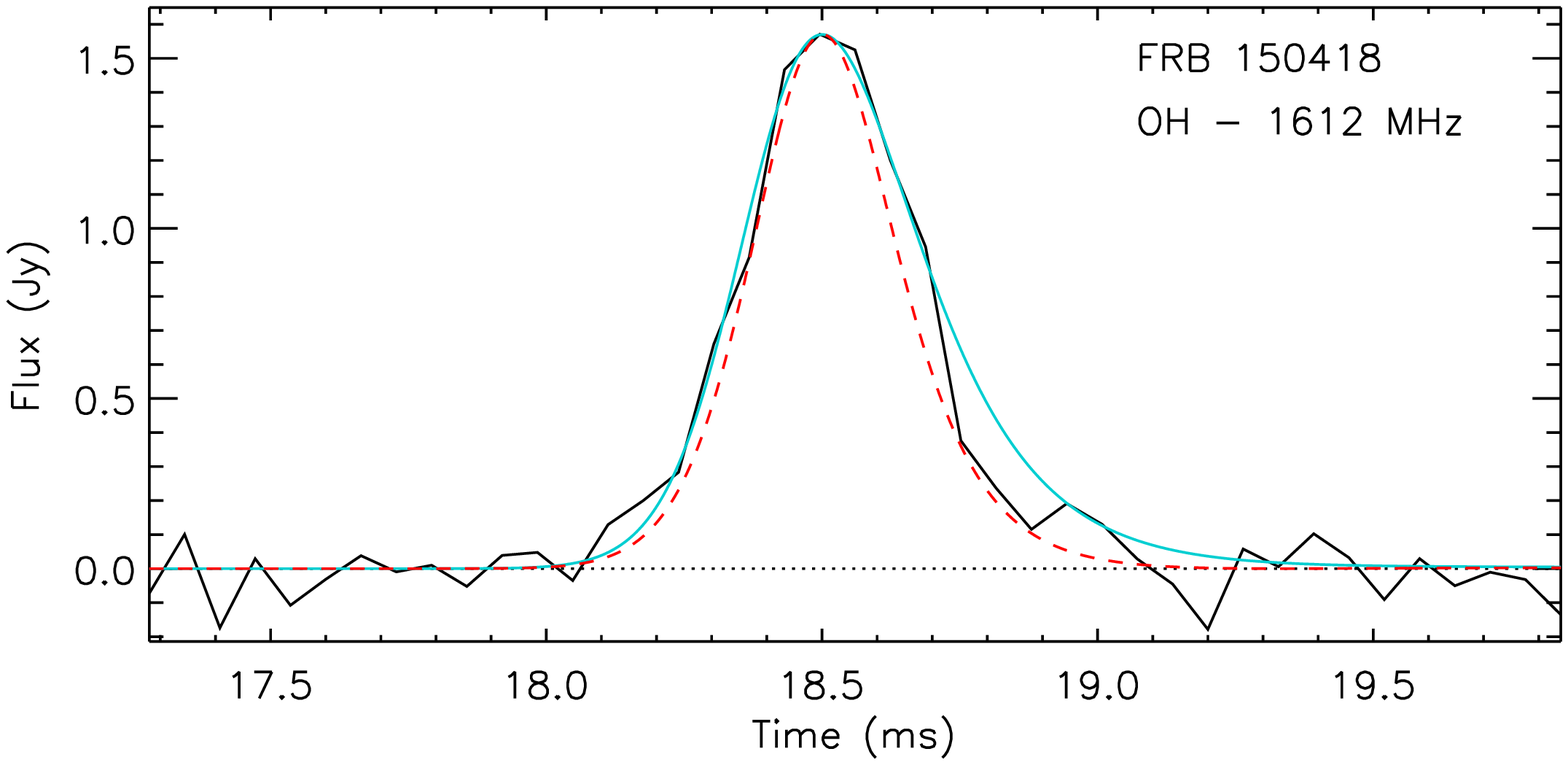}
        \caption{Same as Fig. \ref{fig:FRB110220} but for FRB 150418 \citep{Keane2016}. The SR model parameters are $\left<T_\mathrm{R}\right>=1.6\,\mu\mathrm{sec}$, $\sigma_{T_\mathrm{R}}=0.008\left<T_\mathrm{R}\right>$, and $T^\prime=500\left<T_\mathrm{R}\right>$. The mean inverted column density resulting from the fit is $\left<nL\right>=1.2\times10^{15}$ cm$^{-2}$, which for $\left<n\right>=0.1$ cm$^{-3}$ yields $\left<L\right>=1.2\times10^{16}$ cm.}
        \label{fig:FRB150418}
    \end{figure}
 \end{center}

We present in Figs. \ref{fig:FRB110220} and \ref{fig:FRB150418} SR models we adapted to FRB 110220 \citep{Thornton2013} and FRB 150418 \citep{Keane2016}, respectively, using the large sample SR formalism detailed in Appendix \ref{sec:modelling}. Although the fits to the data are very good, too much emphasis should not be put on this aspect of the results. Rather, these analyses are presented with the intention of demonstrating what sort of pulses SR systems can produce (e.g., multiple peaks) and their appropriateness to FRB signals. In that respect, FRB 110220 is a good example of a profile exhibiting a ``shoulder" and an exponential tail, while FRB 150418 is more symmetric in appearance. Furthermore, these FRBs have two of the highest signal-to-noise ratios from the known detections to date.

In both figures the black and cyan solid curves trace, respectively, the data and the resulting fits, while the broken red curves are those stemming from a single SR sample characterized by the mean characteristic timescale $\left<T_\mathrm{R}\right>$. For FRB 110220 the model parameters are $\left<T_\mathrm{R}\right>=20\,\mu\mathrm{sec}$, $\sigma_{T_\mathrm{R}}=0.07\left<T_\mathrm{R}\right>$, and $T^\prime=850\left<T_\mathrm{R}\right>$, while for FRB 150418 $\left<T_\mathrm{R}\right>=1.6\,\mu\mathrm{sec}$, $\sigma_{T_\mathrm{R}}=0.008\left<T_\mathrm{R}\right>$, and $T^\prime=500\left<T_\mathrm{R}\right>$. The SR analyses provide us with estimates of the mean inverted column density for both FRBs, namely $\left<nL\right>=9.5\times10^{13}$ cm$^{-2}$ for FRB 110220 and $\left<nL\right>=1.2\times10^{15}$ cm$^{-2}$ for FRB 150418. Assuming a mean inverted density of $\left<n\right>=0.1$ cm$^{-3}$, as expected for OH masers, yields $\left<L\right>=9.5\times10^{14}$ cm and $1.2\times10^{16}$ cm for FRB 110220 and FRB 150418, respectively. The spectral width $\Delta\nu$ associated with an FRB is set by the smallest temporal structure in its intensity curve (e.g., the first lobe of the red broken curve in Fig. \ref{fig:FRB110220}) and scales as $\Delta\nu\propto T_\mathrm{R}^{-1}$. We find $\Delta\nu\simeq 300$ Hz for FRB 110220 and $\Delta\nu\simeq 1400$ Hz for FRB 150418; narrower FRBs would have correspondingly larger bandwidth. These spectral widths would approximately correspond to the correlated frequency scale expected for these FRBs \citep{Ravi2016} and are consistent with having dephasing effects dominated by Doppler motions in the SR molecular populations (see Appendix \ref{sec:spectrum}). The presence of multiple peaks in the single SR sample of Fig. \ref{fig:FRB110220} (red broken curve) for FRB 110220 is due to successive emission/absorption episodes within the molecular populations. A phenomenon known as ``ringing effect'' (see Appendix \ref{sec:large-sample}). 

We note that for a given FRB the data dictate the value for $\left<T_\mathrm{R}\right>$ but the ensuing estimate for $\left<nL\right>$ depends on the spectral line chosen for the fit. For example, a stronger line with a spontaneous emission timescale $\tau_\mathrm{sp}$ an order of magnitude smaller would lead to an inverted column density also ten times smaller (see Eq. \ref{TR}). Similarly, the chosen value for the mean inverted density $\left<n\right>$ has a direct impact on the length of the SR sample $\left<L\right>$. Still, our analyses imply a typical SR sample size on the order of 100 AU to 1000 AU, which is a reasonable size for regions harboring inverted molecular populations. For example, the amplification length for megamasers in active galactic nuclei (AGN) environments is expected to be on the order of 1000 AU. Likewise, our column density estimates are also consistent with those for OH starburst megamasers \citep{Gray2012}. 

The SR models for the two FRBs also reveal them to be of small cross-section, with radii of approximately 750 km for FRB 110220 and 2600 km for FRB 150418. Moreover, despite their small extent the SR samples are extremely powerful, in view of the coherent nature of their radiation and the large number of molecules they contain (i.e., $N\sim10^{30}$ and $\sim10^{32}$ for FRB 110220 and FRB 150418, respectively). More precisely, we calculate that, at a distance of 1 Gpc, one single SR sample would exhibit an integrated flux of $\sim10^{-31}\;\mathrm{W\,m}^{-2}$ for FRB 110220 and $\sim10^{-27}\;\mathrm{W\,m}^{-2}$ for FRB 150418. Since the regions harboring inverted molecular populations are expected to be much wider than the aforementioned one-SR sample cross-sections (e.g., the spot size of maser regions are found to be on the order of 10 AU or more), it follows that a staggeringly large number of SR samples could radiate simultaneously and easily match the detected flux associated with FRB signals.  

We therefore find that SR systems are seemingly capable of reproducing the observed timescales, as well as intensity levels and profiles of FRBs through the emission of coherent radiation over relatively small spatial extent. We also note that the detection of one FRB at several frequencies (i.e, FRB 121102 \citep{Scholz2016,Spitler2014,Spitler2016,Law2017,Gajjar2017}) favors the notion that FRBs are due to radiation from spectral lines. The excitation of various transitions, probably from different molecules, is consistent with the existence of FRBs in different bands separated by frequency gaps devoid of signals. Furthermore, if SR were to occur in only one non-degenerate transition, then this would naturally result in a signal exhibiting high levels of polarization, as is sometimes observed \citep{Petroff2015,Masui2015,Ravi2016}. On the other hand, if two or more degenerate spectral lines exhibiting different polarization modes were to simultaneously sustain SR radiation, then polarization levels could be greatly reduced or even cancelled, depending on the relative amounts of radiation emanating from each line (as is the case for molecular lines in general \citep{Goldreich1981}). 
Interestingly, in the case of FRB 121102 at a known redshift $z=0.193$ \citep{Tendulkar2017} the bursts detected at $\sim1.4$ GHz point to the OH lines at 1.6-1.7 GHz, those in the 2.5-3.5 GHz band to the CH lines at $\sim3.3$ GHz, while the 4-8 GHz detections suggest the OH 6 GHz, CH$_3$OH 6.7 and HC$_3$N 9.1 GHz lines (see Table 2.1 in \citealt{Gray2012}). Although not all lines known to exhibit population inversion must necessarily lead to SR, it will be interesting to find out if future observational studies will, for example, detect bursts around 4 GHz, which could be associated to the OH 4.7 GHz and/or H$_2$CO 4.8 GHz maser lines. 

Furthermore, the existence of a group of lines clustered in frequency could result in interesting observable effects in FRB detections within a specific spectral band. For example, let us assume that all four OH $^2\Pi_{3/2}$ transitions (i.e., the main lines at 1665 MHz and 1667 MHz, and the 1612 MHz and 1720 MHz satellite lines) were simultaneously pumped beyond their respective critical inverted column density thresholds of Eq. \ref{nL_crit} and therefore all partaking in some SR event. As mentioned above, for FRB 121102 these signals would be detected in the $\sim1.4$ GHz frequency band. However, because of their different parameters, i.e., their wavelengths and spontaneous emission timescales ($\tau_\mathrm{sp}$), each line will have a different value for $T_\mathrm{R}$ and $\tau_\mathrm{D}$, from Eqs. \ref{TR} and \ref{tau_D}, respectively. It therefore follows that we should expect the four corresponding SR bursts to happen at different times and last for varying durations. 

\begin{center}
   \begin{figure}
        \includegraphics[width=\columnwidth]{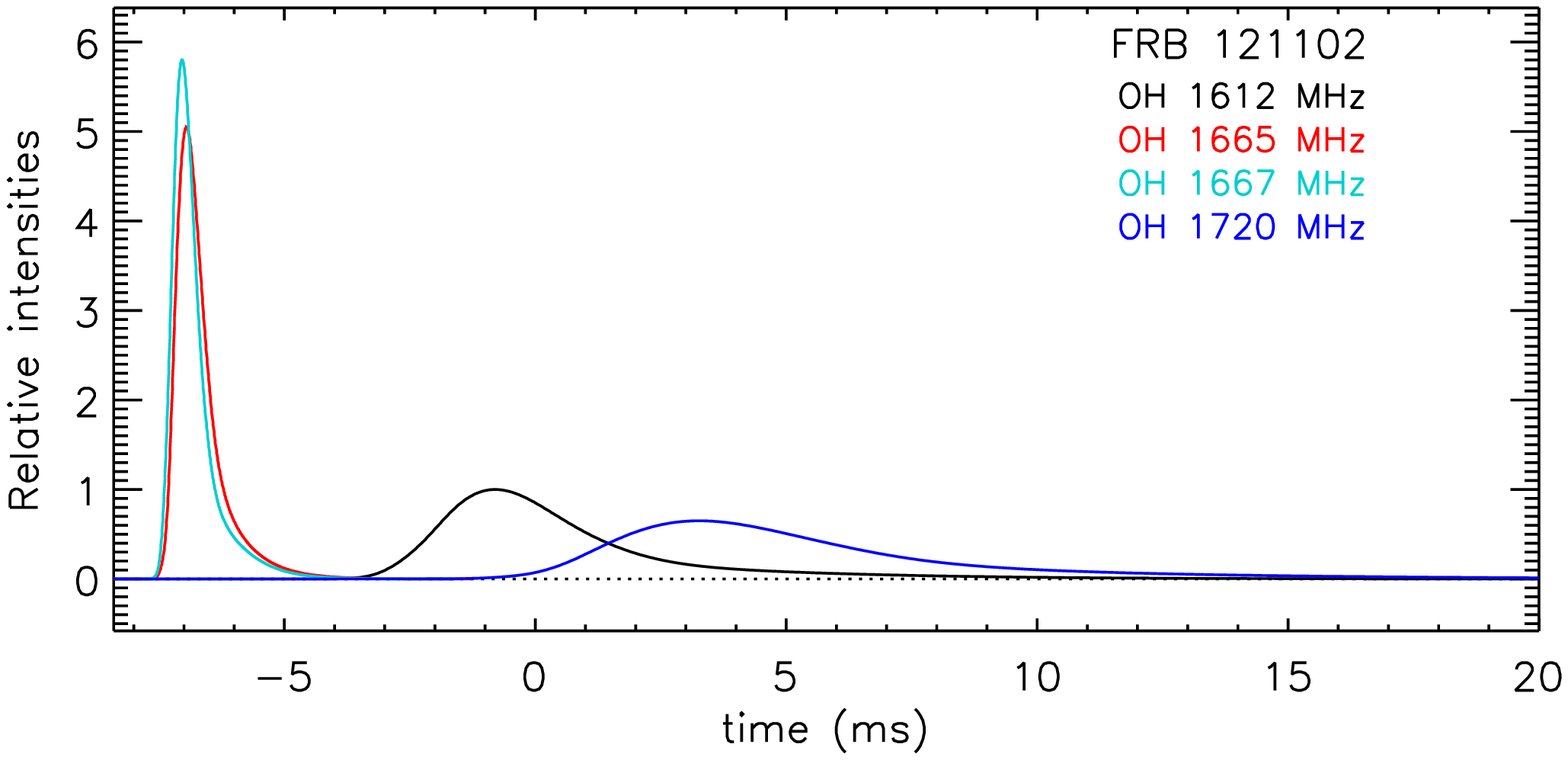}
        \caption{Possible SR responses from the four OH $^2\Pi_{3/2}$ transitions when triggered simultaneously (at $-8.4$ ms). Because of their different values of SR characteristic timescale $T_\mathrm{R}$ and time delay $\tau_\mathrm{D}$, the four SR bursts will happen at different times and exhibit different durations. Given their similarities in wavelength and spontaneous emission timescale, the two main lines at 1665 MHz and 1667 MHz would probably be indiscernible and appear as one burst in observations. Such a system would then reveal three distinct FRBs within a few milliseconds, each covering slightly shifted frequency band. Note that different parameters would alter the relative timescales and intensities of the four bursts (see text for more details).}
        \label{fig:FRB121102}
    \end{figure}
 \end{center}

We show a possible scenario in Fig. \ref{fig:FRB121102} for FRB 121102. To do so we have arbitrarily modelled the 1612 MHz satellite line to the discovery burst of FRB 121102 \citep{Spitler2014} in order to set the timescale $\left<T_\mathrm{R}\right>$, and further assumed that all four OH $^2\Pi_{3/2}$ lines share the same value $\left<nL\right>$ for their inverted column densities. Although the latter condition is unlikely to be realized in any given FRB source, it could be relaxed without altering the essence of our results (see below). As can be seen from the figure, although the four transitions were simultaneously triggered (at $-8.4$ ms), as expected their respective SR bursts do not appear at the same time and exhibit different durations. The stronger 1665 MHz and 1667 MHz main lines have more powerful and narrower bursts, and appear before the other two because of their shorter $\left<T_\mathrm{R}\right>$ and $\left<\tau_\mathrm{D}\right>$ timescales (both parameters scale linearly with $\tau_\mathrm{sp}$, which is approximately $7.8\times10^{10}\,\mathrm{s}$, $1.4\times10^{10}\,\mathrm{s}$, $1.3\times10^{10}\,\mathrm{s}$ and $1.1\times10^{11}\,\mathrm{s}$ for the 1612 MHz, 1665 MHz, 1667 MHz and 1720 MHz transitions, respectively; see Eqs. \ref{TR} and \ref{tau_D}). But given their similarities in wavelength and spontaneous emission timescale, these two spectral lines would probably be indiscernible and appear as one burst in observations. Such system would then reveal three distinct FRBs within a few milliseconds, each covering slightly shifted frequency bands.    

We stress that some of our choices leading to the numerical results presented in Fig. \ref{fig:FRB121102} are arbitrary and other equally valid assumptions could alter the profiles and timescales of the FRBs appearing in the figure. For example, choosing another transition (i.e., other than the 1612 MHz line) to model the discovery burst would change the timescales for all bursts, while different values for the inverted column density for the different lines would also alter their relative intensities. Still, the detection of distinct FRB signals separated by a few milliseconds over slightly shifted frequency band would provide evidence of SR for FRB 121102.

The existence of a critical threshold to be reached for the onset of SR, see Eq. \ref{nL_crit}, can also naturally lead to recurring or repeating FRB signals. For example, one can imagine a situation where the level of inversion is increasing with time through some pumping mechanism until the aforementioned critical value is reached. At that point SR is set-off, an FRB is emitted, while the inverted population is rapidly quenched (assuming the timescale of inversion pumping is significantly longer than $T_\mathrm{R}$). If the pumping mechanism is still at work after the emission of the SR pulse, then the inverted population will be made to increase anew until the critical threshold is once again reached with the ensuing emission of a subsequent SR pulse, and so on. Such scenario has been observed in SR laboratory experiments \citep{Gross1976}, while evidence for repeating SR pulses has been recently uncovered in the ISM \citep{Fujisawa2012,Szymczak2016,Rajabi2017}.     
Our SR model rests on two further requirements. First, the length of the characteristic timescale $T_\mathrm{R}$ (as low as $\sim1\;\mu\mathrm{sec}$) needed to match the duration of FRBs implies that the radiation propagation time through the SR sample's length (as much as $\sim1000\;\mathrm{AU}$) is greater than $T_\mathrm{R}$. This puts a constraint on the type of pumping needed to establish the population inversion. More precisely, a uniform or transverse inversion that would be established more or less instantaneously throughout the SR sample could not lead to the short pulses observed. This is because regions within the sample separated by a distance $d>cT_\mathrm{R}$ would then independently emit photons before interacting with each other, causing the inverted region to break out into an ensemble of smaller SR samples (of longer $T_\mathrm{R}$ values). This is the well-known Arecchi-Courtens condition for SR \citep{Arecchi1970,Gross1982}. However, this condition does not apply for radiation-inverted systems through a swept or longitudinal excitation because molecules are only inverted when the pumping ``wave" propagating through the medium arrives at their position. Independent SR sub-samples cannot form under these conditions and the cooperation length can theoretically be infinite \citep{MacGillivray1976,Gross1982}. This type of swept pumping by radiation is implicit to our model, permitting either single or repeating pulses depending on the nature of the external pump.

Finally, as with any model seeking to reproduce observed FRB signals, our SR analysis must account for their large spectral extent. If the spectral widths associated with the SR intensity curves such as those presented in Figs. \ref{fig:FRB110220} and \ref{fig:FRB150418} could well account for the narrow correlation scales suspected to underlie FRB spectra \citep{Ravi2016}, the tens to hundreds of MHz measured widths must be due to environmental conditions. The regions harboring SR-induced FRBs must therefore be permeated with gas exhibiting motions covering the large (relativistic) velocity range needed to produce the aforementioned spectral widths at the observed frequencies. For this, one may consider, as an example, a scaled up version of the AGN associated with luminous infrared galaxies, where beamed emission from megamasers and gigamasers is detected \citep{Booth1998,Lockett2008}.  Velocities ranging from a few hundreds to thousands of $\mathrm{km\,s}^{-1}$ have been observed in such environments \citep{Baan2001,Lo2005}, and we note that a persistent radio source believed to be an AGN has been co-localized to FRB 121102 \citep{Chatterjee2017,Tendulkar2017}. Furthermore, in past surveys of roughly 200 hours of telescope time at the Arecibo Observatory, about 300 (extragalactic) OH megamasers were observed associated to AGN, suggesting velocity coherence in large radiating gas clusters is ubiquitous near the centers of galaxies \citep{Darling2002}. These OH megamasers have compact regions that can be radiatively pumped where $L\sim1$ parsec and $n\sim0.1\, \mathrm{cm}^{-3}$ \citep{Kylafis1998}. Our SR model for FRBs relies on the existence of population-inverted regions such as those found in these AGN, but with an increased velocity range. The total FRB spectrum would then be composed of a very large number of individual transient SR spectra, each of small width $\sim\Delta\nu$ but uncorrelated among themselves, in a manner consistent with the autocorrelations of FRB spectra \citep{Ravi2016}.

\section{Conclusions}\label{sec:Conclusion}

Our application of Dicke's SR, for which evidence has recently been found in the ISM, reveals that it can account for several characteristics associated to FRBs (e.g., timescales and profiles, intensities, polarization, single or repeating pulses). Our analysis suggests that FRBs could originate from regions similar to those known to harbor masers or megamasers, and result from the coherent radiation emanating from populations of molecules associated with large-scale entangled quantum mechanical states. Using the OH 1612 MHz line for our analysis, we estimate this entanglement to involve as many as $\sim10^{30}$ to $\sim10^{32}$ molecules over distances spanning 100 to 1000 AU.

\section*{Acknowledgements}

We thank E. Keane for sharing his data of FRB 150418, as well as B. Gaensler, S. Tendulkar and C. Law for helpful comments and discussions. M.H.'s research is funded through the Natural Sciences and Engineering Research Council of Canada Discovery Grant RGPIN-2016-04460 and the Western Strategic Support for Research Accelerator Success. 




\bibliographystyle{mnras}
\bibliography{FRB-bib} 




\appendix

\section{Superradiance theory and modelling}\label{sec:modelling}

The material presented in this section (with the exception of Sec. \ref{sec:spectrum}) closely follows the formalism previously developed in the literature (see for example \citealt{Dicke1954,Arecchi1970,Gross1982,Benedict1996,Rajabi2016A,Rajabi2016B}, where more details can be found). 

We model the SR Hamiltonian for a system of $N$ two-level molecules interacting with their common electromagnetic field with 

\begin{equation}
    \hat{H}=\hat{H}_\mathrm{mol}+\hat{H}_\mathrm{rad}+\hat{V} \label{eq:Hamiltonian}
\end{equation}

\noindent where the molecular, radiation field and interaction components of the Hamiltonian are given by

\begin{eqnarray}
    \hat{H}_\mathrm{mol} & = & \hbar\omega_0\sum_{j=1}^N \hat{R}_j^3 \label{eq:Hmol} \\
    \hat{H}_\mathrm{rad} & = & \sum_{\mathbf{k},\zeta}\hbar\omega_k\left(\hat{a}^\dag_{\mathbf{k},\zeta}\hat{a}_{\mathbf{k},\zeta}+\frac{1}{2}\right) \label{eq:Hrad} \\
    \hat{V} & = & -\sum_{j=1}^N\mathbf{\hat{D}}_j\cdot\mathbf{\hat{E}}\left(\mathbf{r}_j\right). \label{eq:V}
\end{eqnarray}

\noindent with, for the electric dipole moment operator of molecule $j$,

\begin{equation}
    \mathbf{\hat{D}}_j = \left(\hat{R}_j^+ +\hat{R}_j^-\right)\mathbf{d}. \label{eq:D}
\end{equation}

\noindent In this system of equations $\hat{R}_j^3$ (where `3' stands for the quantization $z$-axis) and $\hat{R}_j^\pm=\hat{R}_j^x\pm i\hat{R}_j^y$ are pseudo-spin operators, and $\mathbf{d}$ the vector matrix element for the molecular electric dipole moment common to all molecules. The separation between the two molecular energy levels is given by $\hbar\omega_0$, while $\hat{a}^\dag_{\mathbf{k},\zeta}$ and $\hat{a}_{\mathbf{k},\zeta}$ are, respectively, the photon creation and annihilation operators for the radiation mode of wave vector $\mathbf{k}$ and polarization state $\zeta$ (the angular frequency of radiation is $\omega_k=kc$, with $k=\left|\mathbf{k}\right|$). The (radiation) electric field common to all molecules in the sample is given by  

\begin{eqnarray}
   \mathbf{\hat{E}}\left(\mathbf{r}\right) & = & i\sum_{\mathbf{k},\zeta}\boldsymbol{\varepsilon}_{\zeta}\mathcal{E}_\mathbf{k}^{\left(1\right)}\left(e^{i\mathbf{k}\cdot\mathbf{r}}\hat{a}_{\mathbf{k},\zeta}-e^{-i\mathbf{k}\cdot\mathbf{r}}\hat{a}^\dag_{\mathbf{k},\zeta}\right) \label{eq:E(r)} \\
   & = & \mathbf{\hat{E}}^+\left(\mathbf{r}\right) + \mathbf{\hat{E}}^-\left(\mathbf{r}\right) \label{eq:E+-(r)}
\end{eqnarray}

\noindent with $\boldsymbol{\varepsilon}_{\zeta}$ the unit vector defining the polarization state and the one-photon electric field

\begin{equation}
    \mathcal{E}_\mathbf{k}^{\left(1\right)} = \sqrt{\frac{\hbar\omega_k}{2\epsilon_0 \mathcal{V}}}, \label{eq:one-photon}
\end{equation}

\noindent where $\mathcal{V}$ is the quantization volume and $\epsilon_0$ the permittivity of vacuum \citep{Grynberg2010}.

\subsection{The small sample}\label{sec:small-sample}

When the linear dimension of the volume containing the molecules is significantly smaller than the wavelength of radiation, i.e., $L\ll \lambda$, the system operates in the so-called ``small sample'' regime. The problem can then be effectively analyzed using the Schr\"{o}dinger representation where one particular group of molecular eigenstates (i.e., the Dicke states) can be shown to be symmetric under the permutation of any pair of molecules. For example, when $N=3$ these eigenstates are 

\begin{eqnarray}
    \left|3\right> & = & \left|eee\right> \label{eq:|a>} \\
    \left|2\right> & = & \frac{1}{\sqrt{3}}\left(\left|gee\right>+\left|ege\right>+\left|eeg\right>\right) \label{eq:|b>} \\
    \left|1\right> & = & \frac{1}{\sqrt{3}}\left(\left|gge\right>+\left|egg\right>+\left|geg\right>\right) \label{eq:|c>} \\
     \left|0\right> & = & \left|ggg\right> \label{eq:|d>},
\end{eqnarray}

\noindent where $\left|e\right>$ and $\left|g\right>$ are the upper and lower molecular states, respectively. The entangled states $\left|1\right>$ and $\left|2\right>$ result from the interaction within the sample, and are at the heart of the effect responsible for the enhanced radiation intensities and reduced time-scales in SR.

More precisely, in the absence of interaction within an arbitrary sample the $N$ molecules (initially in the upper state $\left|e\right>$) evolve independently from one another and would be expected to each spontaneously emit a photon through the electric dipole transition $\left<e\left|\mathbf{\hat{D}}\cdot\boldsymbol{\varepsilon}_{\zeta}\right|g\right>$ over a time-scale $\tau_\mathrm{sp}=\Gamma^{-1}$ ($\Gamma$ is the Einstein spontaneous emission coefficient for that transition). The total non-coherent radiation intensity as a function of time is thus given by

\begin{equation}
    I_\mathrm{nc}\left(t\right) = N\Gamma\hbar\omega_0 e^{-t/\tau_\mathrm{sp}}, \label{eq:I_nc}
\end{equation}

\noindent where $\Gamma\propto\left|\left<e\left|\mathbf{\hat{D}}\cdot\boldsymbol{\varepsilon}_{\zeta}\right|g\right>\right|^2$ \citep{Grynberg2010}.

In contrast, the molecules in the previous three-molecule SR sample cannot be thought of as independent entities since their entanglement implies that they will evolve as a single unit. Instead of experiencing three independent non-coherent spontaneous emissions the SR system will go through the coherent radiative cascade $\left|3\right>\rightarrow\left|2\right>\rightarrow\left|1\right>\rightarrow\left|0\right>$, while emitting a photon at each transition. Furthermore, because the electric dipole moment of the sample is now

\begin{equation}
    \mathbf{\hat{D}}_\mathrm{SR} = \sum_{j=1}^3 \mathbf{\hat{D}}_j \label{eq:d_sr}
\end{equation}

\noindent the transition rates $\gamma_{mn}\propto\left|\left<m\left|\mathbf{\hat{D}}_\mathrm{SR}\cdot\boldsymbol{\varepsilon}_{\zeta}\right|n\right>\right|^2$ between two states $\left|m\right>$ and $\left|n\right>$ through the cascades have the following dependencies

\begin{eqnarray}
    \gamma_{32} & \propto & 3\left|\left<e\left|\mathbf{\hat{D}}\cdot\boldsymbol{\varepsilon}_{\zeta}\right|g\right>\right|^2 \label{eq:gamma_32} \\
    \gamma_{21} & \propto & 4\left|\left<e\left|\mathbf{\hat{D}}\cdot\boldsymbol{\varepsilon}_{\zeta}\right|g\right>\right|^2 \label{eq:gamma_21} \\
    \gamma_{10} & \propto & 3\left|\left<e\left|\mathbf{\hat{D}}\cdot\boldsymbol{\varepsilon}_{\zeta}\right|g\right>\right|^2, \label{eq:gamma10}
\end{eqnarray}

\noindent from Eqs. \ref{eq:|a>}-\ref{eq:|d>} and \ref{eq:d_sr}.

Although the first transition rate imply that the radiation intensity at $t=0$ of the SR sample will be the same as that of the non-coherent sample (compare Eqs. \ref{eq:I_nc} with $N=3$ and \ref{eq:gamma_32}), the middle transition rate is seen to be enhanced in comparison. This increase in transition rate becomes more pronounced as the number of molecules $N$ in the sample augments. It is found that the maximum transition rate always peaks in the middle of the SR cascade and can be shown to be (for $N$ even; \citealt{Dicke1954})

\begin{eqnarray}
    \gamma_\mathrm{max} & \propto & \frac{N}{2}\left(\frac{N}{2}+1\right)\left|\left<e\left|\mathbf{\hat{D}}\cdot\boldsymbol{\varepsilon}_{\zeta}\right|g\right>\right|^2 \nonumber \\
    & \propto & N^2\left|\left<e\left|\mathbf{\hat{D}}\cdot\boldsymbol{\varepsilon}_{\zeta}\right|g\right>\right|^2. \label{eq:gamma_max}
\end{eqnarray}

The transitions rates are therefore drastically increased from their non-coherent counterpart (which scales as $\left|\left<e\left|\mathbf{\hat{D}}\cdot\boldsymbol{\varepsilon}_{\zeta}\right|g\right>\right|^2$) whenever $N\gg1$. The same is true for the intensity of radiation, which can be shown to be \citep{Benedict1996}

\begin{equation}
    I_\mathrm{SR} = \frac{N^2}{4}\Gamma\hbar\omega_0\cosh^{-2}\left[\frac{1}{2}N\Gamma\left(t-t_\mathrm{D}\right)\right]. \label{eq:I_SR}
\end{equation}

\noindent The SR small sample is thus seen to emit a powerful pulse of radiation over a drastically shorter characteristic time-scale on the order of

\begin{equation}
    T_\mathrm{R} = \frac{\tau_\mathrm{sp}}{N} \label{eq:T_R-small} 
 \end{equation}
 
\noindent after a  ``delay time''

\begin{equation}
    t_\mathrm{D} = T_\mathrm{R}\ln\left(N\right). \label{eq:t_D}
\end{equation}

\noindent This delay before emission is the time interval needed to establish coherence between the individual molecular electric dipoles throughout the sample. Most importantly, a comparison of Eqs. \ref{eq:I_nc} and \ref{eq:I_SR} reveals that $I_\mathrm{nc}$ scales linearly with $N$ while $I_\mathrm{SR}\propto N^2$, hence the term ``superradiance'' \citep{Dicke1954}.  

Despite the significant increase in radiation intensity in a small sample, in astrophysical environments dephasing and non-coherent relaxation effects (e.g., collisions) are likely to happen on a shorter time-scale than $t_\mathrm{D}$ and inhibit the onset of SR by destroying coherence within the sample \citep{Rajabi2016A,Rajabi2016B}. It is therefore necessary to consider samples of larger dimensions where a greater number of molecules can interact (i.e., to bring a decrease in $T_\mathrm{R}$ and $t_\mathrm{D}$).   

\subsection{The large sample}\label{sec:large-sample}

Although the physics behind SR is perhaps more transparent in the analysis of the small sample, SR can also take place in a so-called ``large sample'' for which $L\gg\lambda$. As phase coherence cannot be preserved simultaneously along all directions whenever some anisotropy is present in the medium under consideration, such samples will tend to have highly elongated profiles and for this reason a thin cylinder of length $L$ is most convenient for the analysis. In what follows the cylindrical axis of symmetry is taken to be aligned with the $z$-direction and, moreover, we consider a one-dimensional model (i.e., with dependencies on only $z$ and $t$).

As the number of molecules increases to very large numbers, the Schr\"{o}dinger representation loses much of its appeal for the SR problem in view of the large number of observables entering the formalism. Also, propagation effects become important in large samples and are better handled through the Heisenberg representation, which we now adopt \citep{Gross1982}. Within this framework  the evolution of the system is tied to that of observables related to the pseudo-spin and and electric field appearing in the Hamiltonian (see Eqs. \ref{eq:Hamiltonian}-\ref{eq:V}). More precisely, the following operators are defined for the molecular polarization and (half the) inverted population density

\begin{eqnarray}
    \mathbf{\hat{P}}^\pm\left(z\right) & = & \mathbf{d}\sum_{j=1}^N\delta\left(z-z_j\right)\hat{R}_j^\pm \label{eq:P} \\
    \hat{\mathbb{N}}\left(z\right) & = & \sum_{j=1}^N\delta\left(z-z_j\right)\hat{R}_j^3. \label{eq:N}
\end{eqnarray}

The relevant equations for the time evolution are then obtained through the Heisenberg equation

\begin{equation}
    \frac{d\hat{X}}{dt}=\frac{1}{i\hbar}\left[\hat{X},\hat{H}\right], \label{eq:Heisenberg}
\end{equation}

\noindent where $\hat{X}$ stands for $\mathbf{\hat{P}}^\pm$, $\hat{\mathbb{N}}$ and $\mathbf{\hat{E}}^\pm$ (note that Eq. \ref{eq:Heisenberg} must be applied twice for the electric field). The resulting expressions (i.e., the so-called Bloch-Maxwell equations) are further simplified by tracking the time evolution using the retarded time $\tau=t-z/c$, and through the slowly-varying-envelope-approximation (SVEA) where the polarization and electric field, at resonance (i.e., when $\omega_k=\omega_0$), take the following form

\begin{eqnarray}
     \mathbf{\hat{P}}^\pm\left(z,\tau\right) & = & \hat{P}_0^\pm\left(z,\tau\right)e^{\pm i\omega_0\tau}\boldsymbol{\varepsilon}_d \label{eq:P_svea} \\
    \mathbf{\hat{E}}^\pm\left(z,\tau\right) & = & \hat{E}_0^\pm\left(z,\tau\right)e^{\mp i\omega_0\tau}\boldsymbol{\varepsilon}_d \label{eq:E_svea}
\end{eqnarray}

\noindent with $\boldsymbol{\varepsilon}_d=\mathbf{d}/\left|\mathbf{d}\right|$, and significant variations in the envelopes $\hat{P}_0^\pm\left(z,\tau\right)$ and $\hat{E}_0^\pm\left(z,\tau\right)$ \ are assumed to happen on time- and length-scales much greater than $1/\omega_0$ and $1/k$, respectively.

Finally, the Bloch-Maxwell equations can be augmented by adding phenomenological non-coherent terms for depolarization, as well as for inverted population relaxation and pumping. The resulting equations are

\begin{eqnarray}
     \frac{\partial\hat{\mathbb{N}}}{\partial\tau} & = & \frac{i}{\hbar}\left(\hat{P}_0^+\hat{E}_0^+-\hat{E}_0^-\hat{P}_0^-\right)-\frac{\hat{\mathbb{N}}}{T_1}+\Lambda_\mathbb{N} \label{eq:dN/dt} \\     \frac{\partial\hat{P}_0^+}{\partial\tau} & = & \frac{2id^2}{\hbar}\hat{E}_0^-\hat{\mathbb{N}}-\frac{\hat{P}_0^+}{T_2} \label{eq:dP/dt} \\
    \frac{\partial\hat{E}_0^+}{\partial z} & = & \frac{i\omega}{2\epsilon_0c}\hat{P}_0^-. \label{eq:dE/dt}
\end{eqnarray}

\noindent The terms $-\hat{\mathbb{N}}/T_1$ and $\Lambda_\mathbb{N}$ in Eq. \ref{eq:dN/dt} are for the non-coherent inverted population relaxation (acting over a characteristic time-scale $T_1$) and pumping, respectively, while in Eq. \ref{eq:dP/dt} $-\hat{P}_0^+/T_2$ causes depolarization (over a time-scale $T_2$).  

The large sample problem requires the simultaneous numerical solution of Eqs. \ref{eq:dN/dt}-\ref{eq:dE/dt} for $\hat{\mathbb{N}}\left(z,\tau\right)$, $\hat{P}_0^+\left(z,\tau\right)$ and $\hat{E}_0^+\left(z,\tau\right)$, from which the radiation intensity is obtained via

\begin{equation}
    I_\mathrm{SR} = \frac{c\epsilon_0}{2}\left|\hat{E}_0^+\right|^2.
\end{equation}

\noindent However, the solution to the problem is significantly simplified when $\Lambda_\mathbb{N}=0$ and only one dephasing/relaxation time-scale $T^\prime=T_1=T_2$ is involved. It can be shown that under these circumstances Eqs. \ref{eq:dN/dt}-\ref{eq:dE/dt} reduce to

\begin{eqnarray}
    \hat{\mathbb{N}}\left(z,\tau\right) & = & \frac{N}{2V}\cos\left(\theta\right)e^{-\tau/T^\prime} \label{eq:N-sg} \\
    \hat{P}_0^+\left(z,\tau\right) & = & \frac{Nd}{2V}\sin\left(\theta\right)e^{-\tau/T^\prime}, \label{eq:P-sg} \\
    \hat{E}_0^+\left(z,\tau\right) & = & \frac{i\hbar}{2d}\frac{\partial\theta}{\partial\tau} \label{eq:E-sg}
\end{eqnarray}

\noindent with $d=\left|\mathbf{d}\right|$ and $V$ the volume containing the large sample. The solution for the Bloch angle $\theta=\theta\left(z,\tau\right)$ is obtained through the so-called sine-Gordon equation

\begin{equation}
    \frac{d^2\theta}{dq^2}+\frac{1}{q}\frac{d\theta}{dq} = \sin\left(\theta\right), \label{eq:sine-Gordon}
\end{equation}

\noindent where 

\begin{equation}
    q = 2\sqrt{\frac{z\tau^\prime}{LT_\mathrm{R}}} \label{eq:q}
\end{equation}

\noindent and $\tau^\prime=T^\prime\left(1-e^{-\tau/T^\prime}\right)$. The initial conditions for Eq. \ref{eq:sine-Gordon} are usually set by considering internal fluctuations (e.g., spontaneous emission) with $\theta\left(\tau=0\right)=2/\sqrt{N}$ and $\partial\theta/\partial\tau\left|_{\tau=0}\right.=0$. The characteristic timescale of SR for the large sample 

\begin{equation}
    T_\mathrm{R} = \tau_\mathrm{sp}\frac{8\pi}{3\lambda^2nL} \label{eq:TR-large}
\end{equation}

\noindent is different from the relation for the small sample (compare with Eq. \ref{eq:T_R-small}), but still inversely proportional to $N$ ($n=N/V$ in Eq. \ref{eq:TR-large}). 

In the top panel of Figure \ref{fig:Efield} we revisit the SR intensity curve associated to FRB 110220 and previously presented in Fig. \ref{fig:FRB110220}, but in a more general manner. That is, the time axis is now for the retarded time $\tau=t-L/c$ at the end-fire of the SR system, where the detected radiation emanates from (i.e., we set $z=L$ in Eq. \ref{eq:q}), and normalized to $\left<T_\mathrm{R}\right>$. This will allow us to apply our results to FRBs of timescales different than those presented earlier in the paper. The vertical intensity is also normalized to $NI_\mathrm{nc}$, where $N$ is, as before, the number of molecules partaking in the SR process and $I_\mathrm{nc}$ is the non-coherent intensity expected from $N$ independent radiators \citep{Rajabi2016B}. Given that our model yields $N\sim10^{30}$ for FRB 110220, this exemplifies how powerful SR bursts are. The black curve shows the intensity averaged over all the cylindrical SR realizations, while the cyan curve corresponds to that of a single SR large sample of mean characteristic timescale $\left<T_\mathrm{R}\right>$. The bottom panel of the figure shows the envelope of the electric field present at the end-fire for that lone SR sample, while Figure \ref{fig:NandP} presents the corresponding inverted population density (top) and polarization envelope (bottom) at the same location. All quantities were computed using Eqs. \ref{eq:N-sg}-\ref{eq:sine-Gordon}, and \ref{eq:I_SR} for the intensity.

The intensity curve in the figure pertaining to the single SR realization (i.e, the cyan curve in the top panel of Figure \ref{fig:Efield}) is a response typical of a large sample with moderate dephasing and non-coherent relaxation (i.e., $T^\prime\gg T_\mathrm{R}$ and at least a few times the delay time $\tau_\mathrm{D}$; see below). That is, we notice the presence of successive bursts of decreasing amplitude with time; the ``ringing effect'' phenomenon previously mentioned in Sec. \ref{sec:discussion}. This effect is related to the fact that the length of the sample greatly exceeds the wavelength of radiation ($L\gg\lambda$), which implies that all molecules within the sample cannot radiate at the same time with the same phase. That is, a group of molecules in the excited state at one location in the sample which radiate its energy in a burst (with the constituent molecules decaying to the ground state) will afterwards absorb the energy radiated by another group of molecules located upstream in the sample, with an ensuing re-emission. This naturally leads to an emission/absorption cycle that is responsible for the appearance of ringing in the intensity curve. The number of bursts is therefore a function of the length $L$ of the sample and the ratio $T^\prime/\tau_\mathrm{D}$. More precisely, the amount of ringing diminishes for decreasing $L$ (because of a lower number of groups of molecules) and/or $T^\prime$ (because of an increase in damping). The broken red intensity curve for the single SR realization for our FRB 150418 model presented in Figure \ref{fig:FRB150418} is a good example for this behaviour, where the decrease in $T^\prime$ relative to $\tau_\mathrm{D}$ (when compared to FRB 110220) limits the number of bursts to one. \citep{Rajabi2016A,Rajabi2016B}.

\subsubsection{Power Spectrum and Spectral Width}\label{sec:spectrum}

\begin{center}
   \begin{figure}
        \includegraphics[width=\columnwidth]{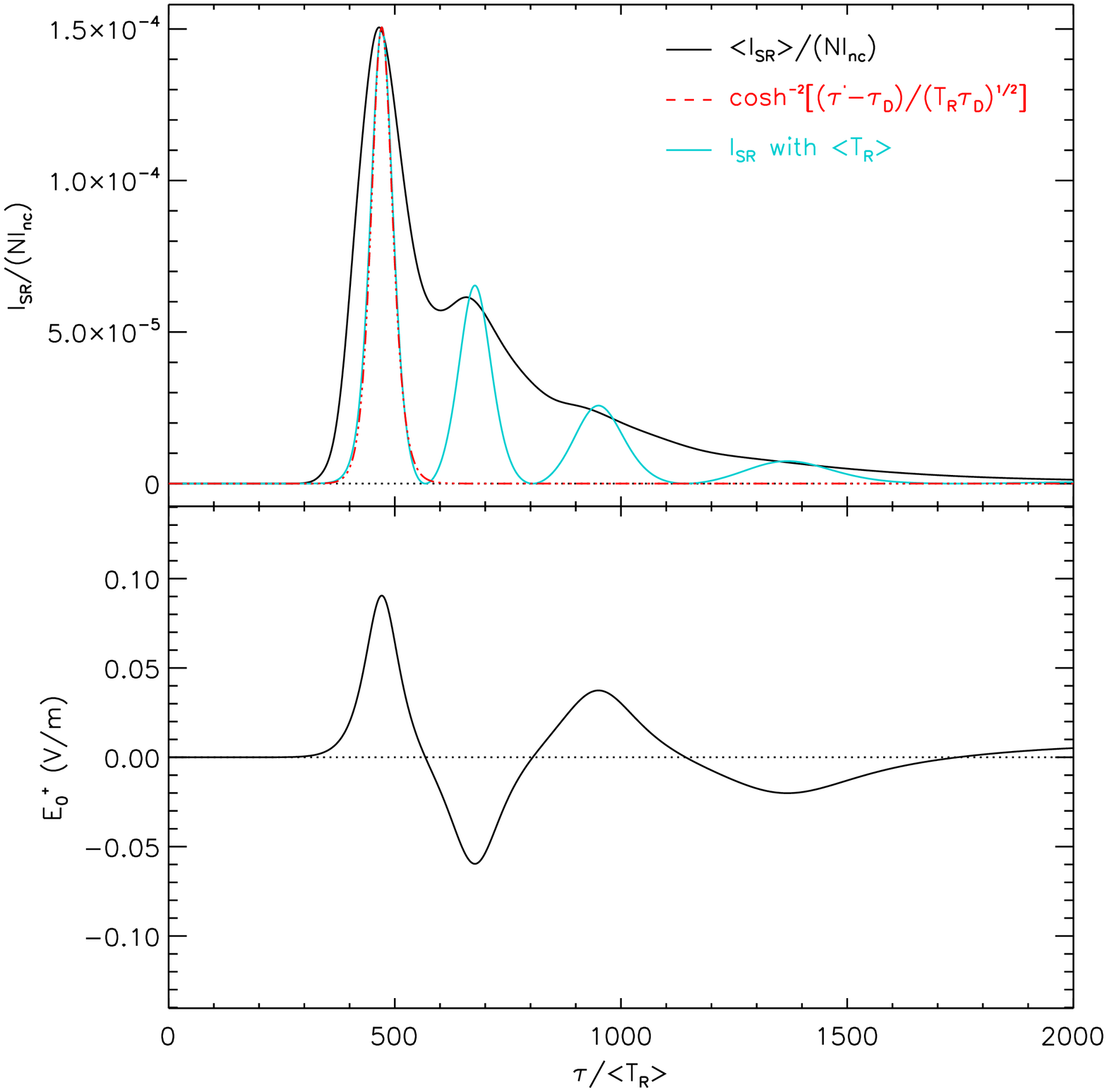}
        \caption{Top: SR intensity model for FRB 110220 (black curve) with the time axis normalized to $\left<T_\mathrm{R}\right>$ and the intensity to $NI_\mathrm{nc}$, where $N$ is the number of molecules partaking in the SR process and $I_\mathrm{nc}$ is the non-coherent intensity expected from $N$ independent radiators. Our model gives $N\sim10^{30}$ for FRB 110220. The cyan curve corresponds to the intensity from one SR sample of mean characteristic timescale $\left<T_\mathrm{R}\right>$. The broken red curve is a functional fit to the first lobe of the single-SR curve (see the text for more details). Bottom: the envelope of the electric field present at the end-fire of the single SR sample.}
        \label{fig:Efield}
    \end{figure}
\end{center}

\begin{center}
   \begin{figure}
        \includegraphics[width=\columnwidth]{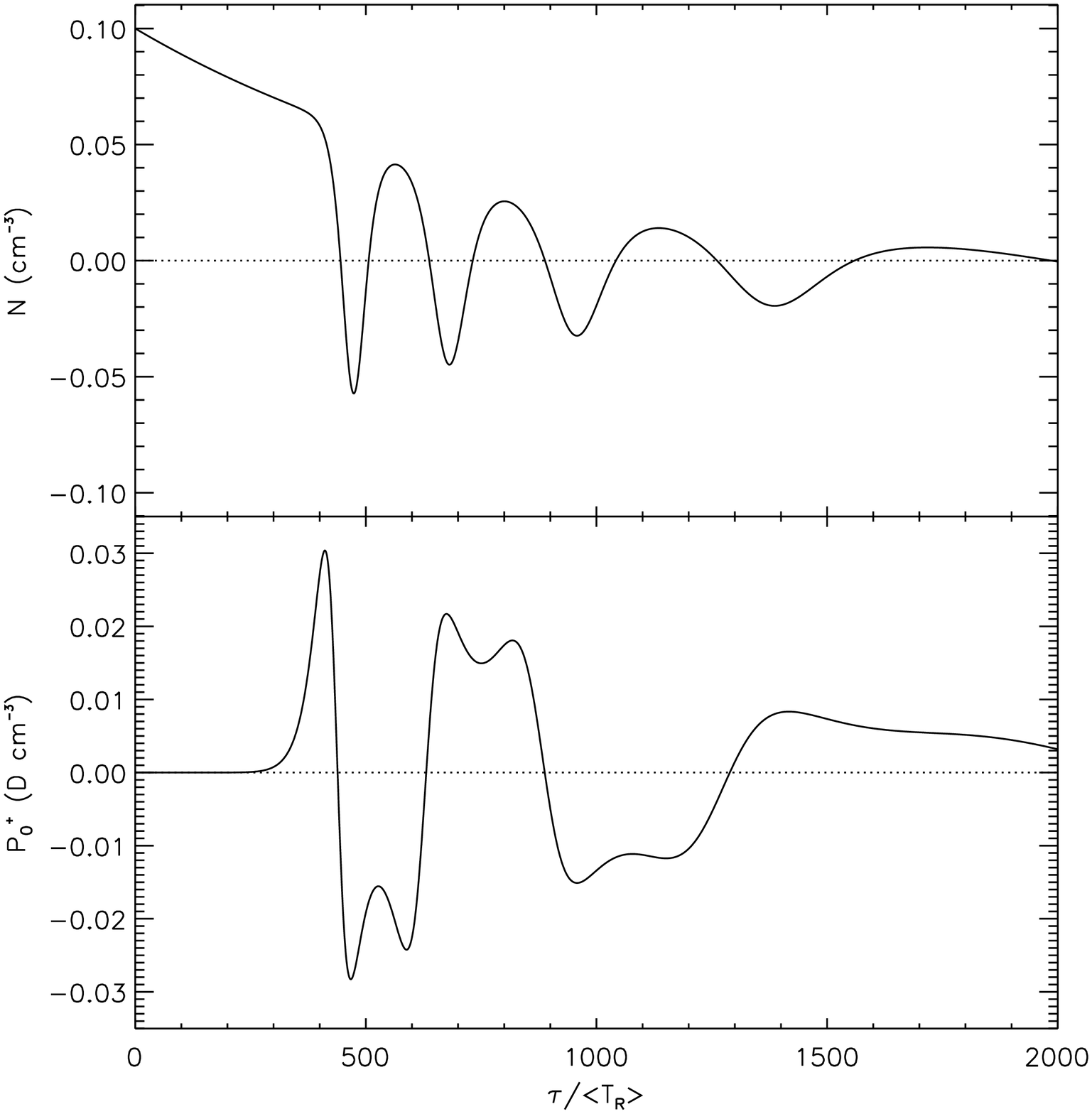}
        \caption{The envelopes of the inverted population density (top) and polarization (bottom) present at the end-fire of the single SR sample for the FRB 110220 model of Figure \ref{fig:Efield}.}
        \label{fig:NandP}
    \end{figure}
\end{center}

The width of the SR spectrum is approximately set by the shortest temporal structure present in the intensity curve. The broken red curve in Fig. \ref{fig:Efield} shows that the first burst of the single-SR curve, which is the shortest timescale, is well fitted by the typical small sample-like SR intensity profile (see Eq. \ref{eq:I_SR})

\begin{equation}
    I_1\propto \cosh^{-2}\left(\frac{\tau^\prime-\tau_\mathrm{D}}{\sqrt{T_\mathrm{R}\tau_\mathrm{D}}}\right), \label{eq:cosh}
\end{equation}

\noindent with the so-called delay time before the emission of the SR burst, defined when no dephasing or non-coherent relaxation are present (i.e., $T^\prime=\infty$), is given by \citep{MacGillivray1976,Gross1982}

\begin{eqnarray}
    \tau_\mathrm{D} & = & \frac{T_\mathrm{R}}{4}\left|\ln\left[\frac{\theta\left(\tau=0\right)}{2\pi}\right]\right|^2 \\
    & = & \frac{T_\mathrm{R}}{4}\left|\ln\left(\pi\sqrt{N}\right)\right|^2 \label{eq:tau_D}.
\end{eqnarray}

Because the width of the first burst is much smaller than the delay time Eq. \ref{eq:cosh} can be rearranged and expressed as a function of $\tau$ (as opposed to $\tau^\prime$) and shown to have a characteristic timescale given by

\begin{equation}
    \Delta\tau=\frac{\sqrt{T_\mathrm{R}\tau_\mathrm{D}}}{1-\tau_\mathrm{D}/T^\prime}.\label{eq:Dtau}
\end{equation}

\noindent Accordingly, the width of the SR spectrum will be given by

\begin{equation}
    \Delta\nu\simeq\frac{1-\tau_\mathrm{D}/T^\prime}{2\pi\sqrt{T_\mathrm{R}\tau_\mathrm{D}}},\label{eq:Dnu}
\end{equation}

\noindent which for FRB 110220 and FRB 150418 yields 275 Hz and 1350 Hz, respectively. 

\begin{center}
   \begin{figure}
        \includegraphics[width=\columnwidth]{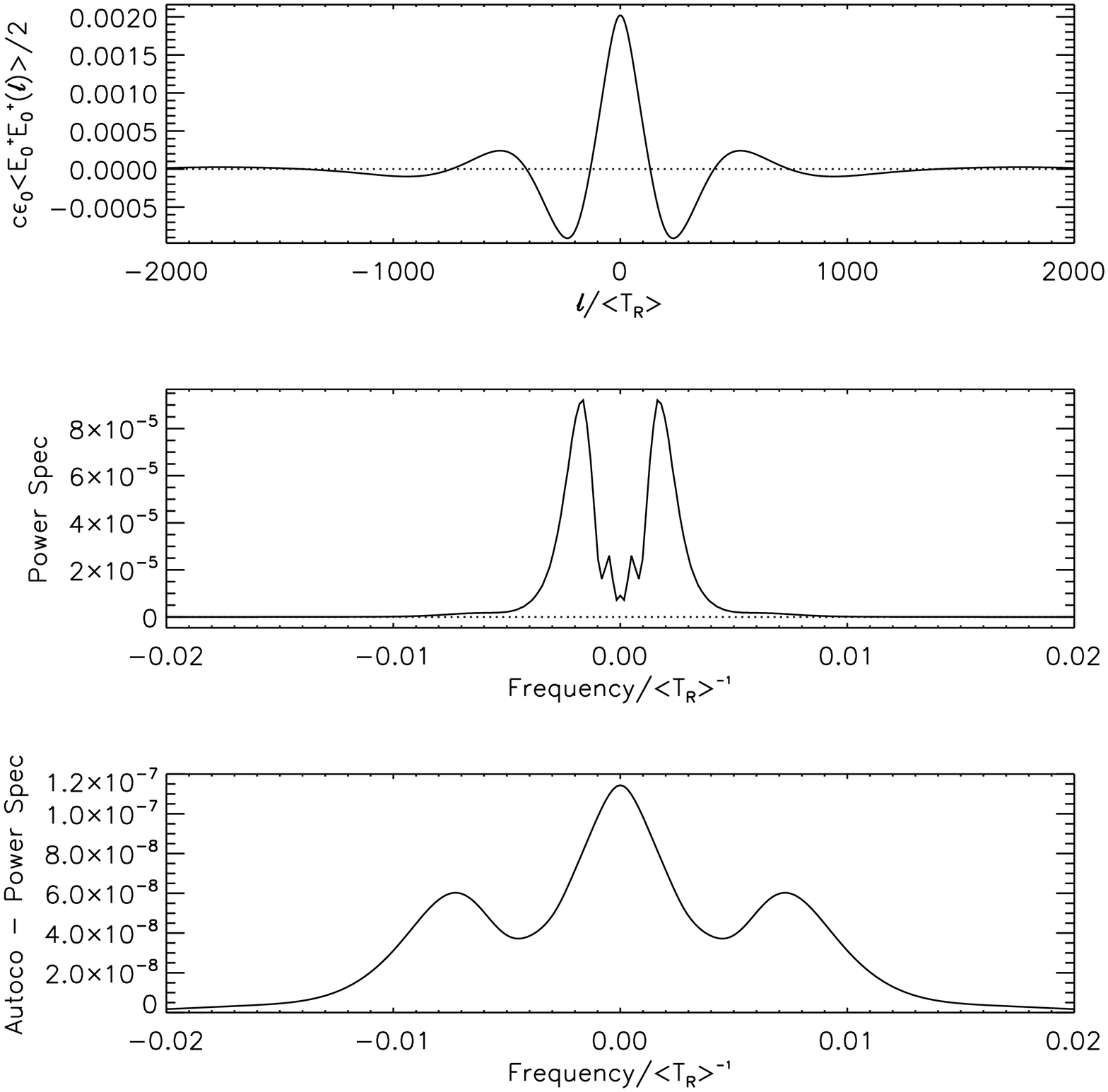}
        \caption{Top: the average of the autocorrelations of the electric field (multiplied by $c\epsilon_0/2$) for the individual SR samples responsible for the total SR curve of Fig. \ref{fig:Efield}, plotted as a function of the time lag $\ell$ normalized to $\left<T_\mathrm{R}\right>$. Middle: the associated power spectrum, consisting of the Fourier transform of the autocorrelation function. The frequency is normalized to $\left<T_\mathrm{R}\right>^{-1}$. Bottom: the autocorrelation of the power spectrum revealing a spectral correlation scale of $\sim 0.01/\left<T_\mathrm{R}\right>^{-1}$, approximately 500 Hz for FRB 110220.}
        \label{fig:pspec}
    \end{figure}
 \end{center}

The top panel of Fig. \ref{fig:pspec} shows the average of the autocorrelations of the electric field (multiplied by $c\epsilon_0/2$) for the individual SR samples responsible for the (black) SR curve of Fig. \ref{fig:Efield}, plotted as a function of the time lag $\ell$ normalized to $\left<T_\mathrm{R}\right>$. The associated power spectrum, consisting of the Fourier transform of the autocorrelation function, is shown in the middle panel of the figure, with the frequency normalized to $\left<T_\mathrm{R}\right>^{-1}$. It can be verified that the width of the spectrum matches well the previous value calculated for FRB 110220 with Eq. \ref{eq:Dnu}. The bottom panel for the autocorrelation of the power spectrum reveals the spectral correlation scale to be $\sim 0.01/\left<T_\mathrm{R}\right>^{-1}$, or approximately 500 Hz for FRB 110220. A FRB with a duration of approximately $300\,\mu\mathrm{sec}$ \citep{Ravi2016,Gajjar2017}, less than a tenth of that of FRB 110220, would thus have a spectrum correlated over almost 10 kHz.   

Finally, we note that we expect ${T^\prime}^{-1}\sim 2\pi \Delta\nu$ in situations where dephasing is dominated by Doppler motions within the inverted population. We therefore obtain from Eq. \ref{eq:Dnu}

\begin{equation}
    T^\prime\sim \tau_\mathrm{D}+\sqrt{T_\mathrm{R}\tau_\mathrm{D}}.\label{Tprime}
\end{equation}

\noindent Our models yield $\tau_\mathrm{D}\simeq320\left<T_\mathrm{R}\right>$ and ${T^\prime}=850\left<T_\mathrm{R}\right>$ for FRB 110220, as well as $\tau_\mathrm{D}\simeq370\left<T_\mathrm{R}\right>$ and ${T^\prime}=500\left<T_\mathrm{R}\right>$ for FRB 150418, which are consistent with this relation. The condition $T^\prime>\tau_\mathrm{D}$ was used to derive Eq. \ref{nL_crit} \citep{Rajabi2016B,Rajabi2017}.

\bsp	
\label{lastpage}
\end{document}